# ON THE FUNCTION $\sum_{k=-\infty}^{\infty} \exp(-\frac{k^2}{s^2})$

**Ioan STURZU***

***Abstract:*** *One is studying numerically and analitically some properties of the function* $y_0(s) = \sum_{k=-\infty}^{\infty} \exp(-\frac{k^2}{s^2})$

The function $y_0(s) = \sum_{k=-\infty}^{\infty} \exp(-\frac{k^2}{s^2})$ is related to the Poisson integral:

$$I(s) = \int_{-\infty}^{\infty} \exp(-\frac{x^2}{s^2})dx = \sqrt{p} \cdot s$$

by the bonding relation: $I(s) < y_0(s) < I(s) + 2$ which can be easy proved using the monotony of the function $\exp(-\frac{x^2}{s^2})$.

However, there are situations [1] when a deeper study of the properties of this function is needed. We shall start by studying the behaviour of $y\ (s)$ in the limiting cases $s \to +0$ and $s \to +\infty$. If $s \to +0$ all the terms in the series go to *0* excepting the term which corresponds to $k=0$. So $y\ (s) \to 1$ when $s \to +0$. When $s \to +\infty$ one can write $y\ (s)$ as a Riemann series for *I(s)*:

$$\frac{1}{s} \cdot y\ \ (s) = \sum_{k=-\infty}^{\infty} \exp(-\frac{k^2}{s^2}) \cdot \frac{\Delta k}{s} \to \int_{-\infty}^{\infty} \exp(-x^2)dx = \sqrt{p}$$

so $y_0(s)$ behaves like *I(s)* when $s \to +\infty$. Let us write $y_0(s)$ in the form: $y_0(s) = \sqrt{p} \cdot s + 1 - e(s)$; than, *e(s)* has the following asymptotic behaviour:

$$\frac{e(s)}{\sqrt{p} \cdot s} \to 1, \text{ when } s \to +0 \ ; \ e(s) \to 1, \text{ when } s \to +\infty .$$

* Physics Department., *Transilvania* University of Braºov.

A reasonable numerical calculus of $y_0(s)$ gives for *e(s)* the plot from Fig. 1.

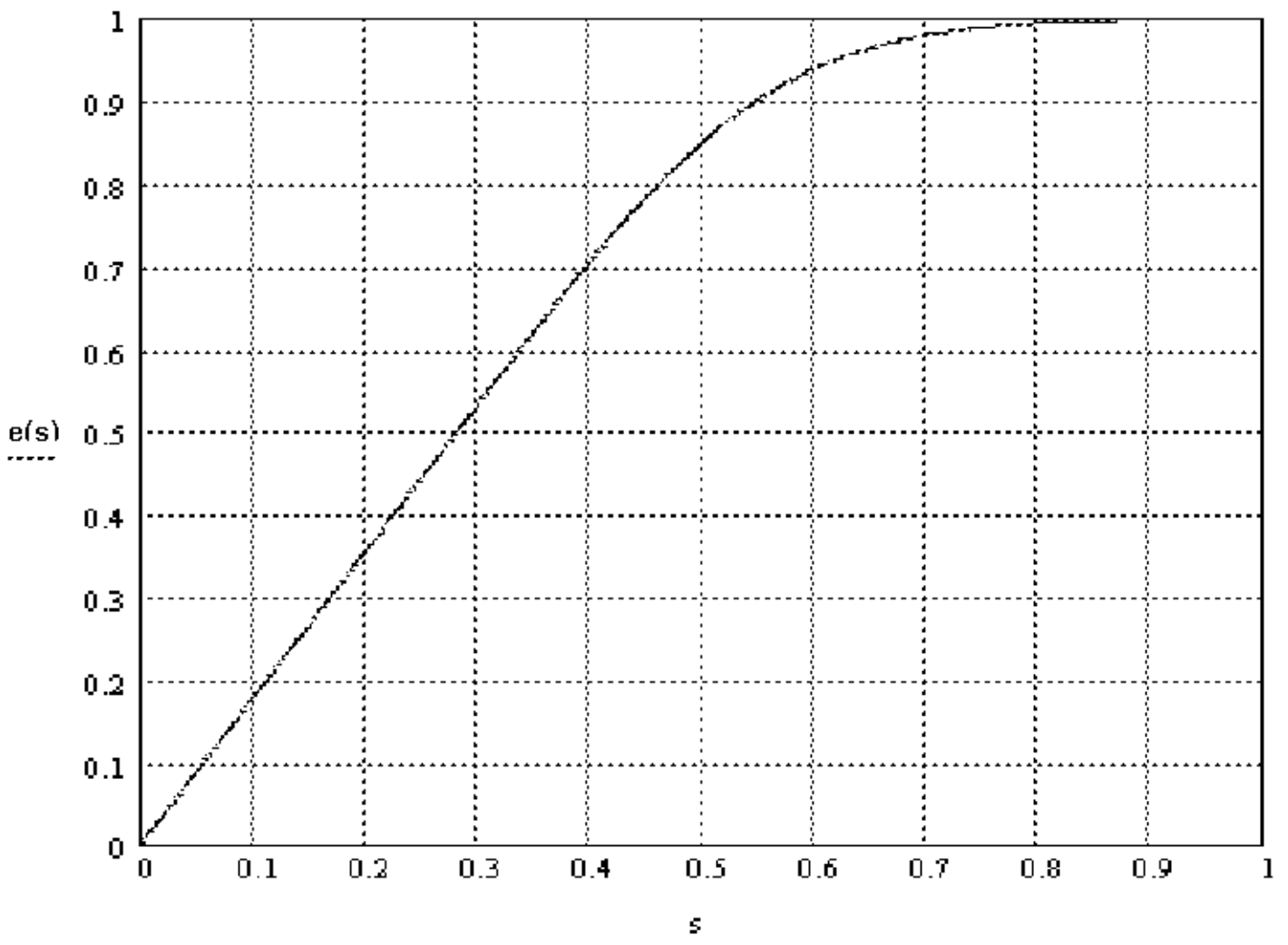

Fig.1. *Numerical calculus for e(s)*

One can easy see that *e(s)* behaves like $\sqrt{p} \cdot s$ for $s \in [0;0.4]$ and is equal to *1* for $s \in [0.8;+\infty]$; so $y_0(s) \approx 1$ for $s \in [0;0.4]$ and $y_0(s) \approx \sqrt{p} \cdot s$ for $s \in [0.8;+\infty]$. For the domain [0.4;0.8] one can obtain a Boltzmann-type fitting:

$$efit(s) = 1.00582 - \frac{0.71664}{1 + \exp(\frac{s - 0.36712}{0.10290})}$$

which is represented in Fig.2.

Let the displaced function of $y_a(s), \quad a \in [0;0.5]$:

$$y_a(s) = \sum_{k=-\infty}^{\infty} \exp(-\frac{(k+a)^2}{s^2})$$

The result of a numerical calculus for $y_{1/2}(s)$ is represented in Fig.3, while the difference $y_0(s) - y_{1/2}(s)$ was numerically interpolated, and yield the formula:

$$diffit(s) = \frac{1}{1 + \exp[\mathrm{sgn}(s - 0.45) \cdot \left| \frac{s - 0.45}{0.088} \right|^{1.29}]}$$

Results are consistent with those obtained from Poisson summation formula [2],[3]:

$$\sum_{k=-\infty}^{\infty} F(\frac{k}{s}) = s \cdot \sum_{k=-\infty}^{\infty} G(s \cdot k)$$

where $G$ is the Fourier Transform of $F$.

For $F(x) = \exp[-(x+a)^2]$, $G(y) = \sqrt{p} \cdot \exp(-p^2 \cdot y^2 - i \cdot a \cdot y)$, so one has the general relation:

$$y_a(s) = \sqrt{p} \cdot s \sum_{k=-\infty}^{\infty} \exp(-p^2 \cdot s^2 \cdot k^2 - i \cdot a \cdot s \cdot k)$$

which goes to:

$$y_0(s) = \sqrt{p} \cdot s \cdot y_0(\frac{1}{p \cdot s})$$

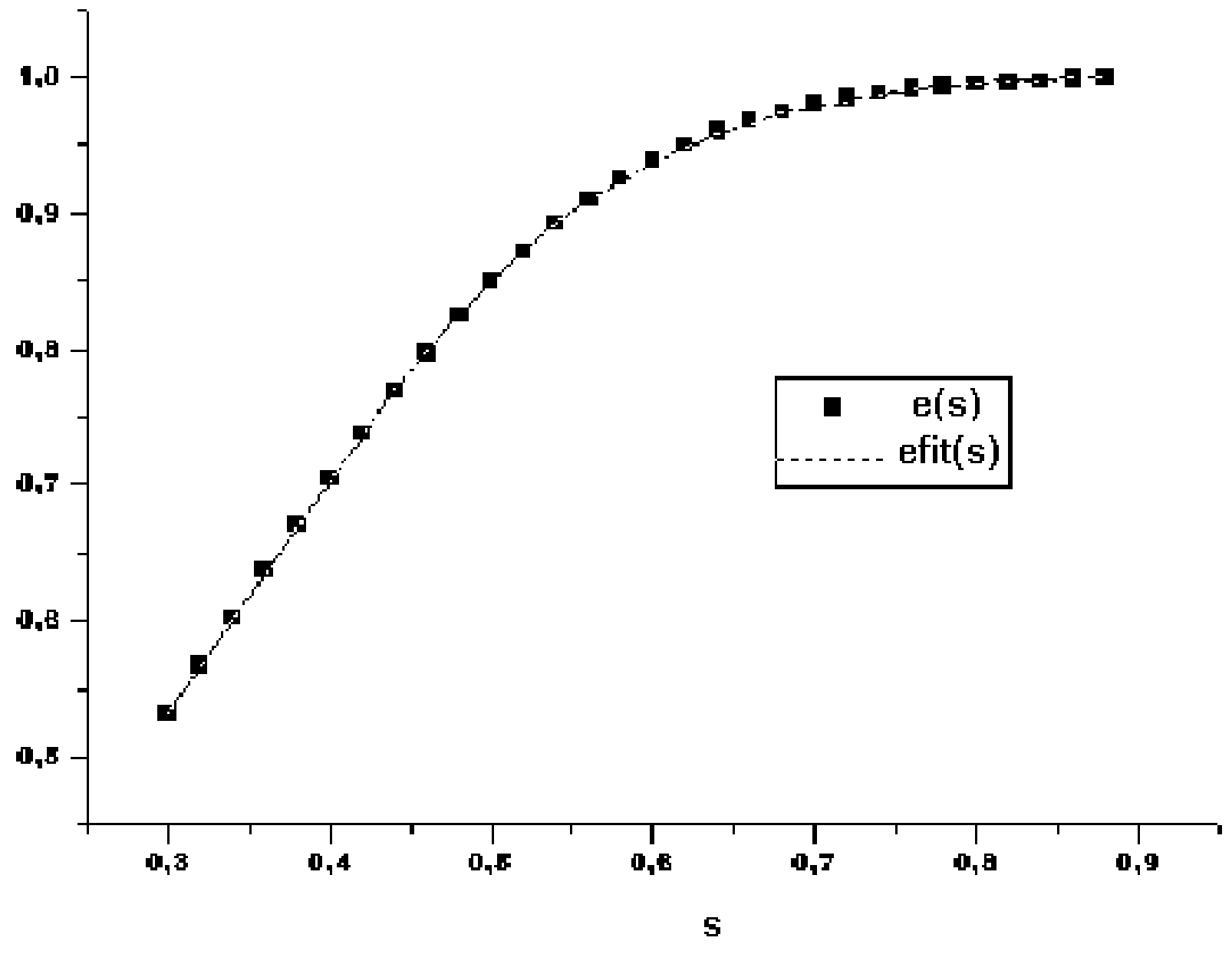

Fig.2. *Numerical interpolation for e(s)*

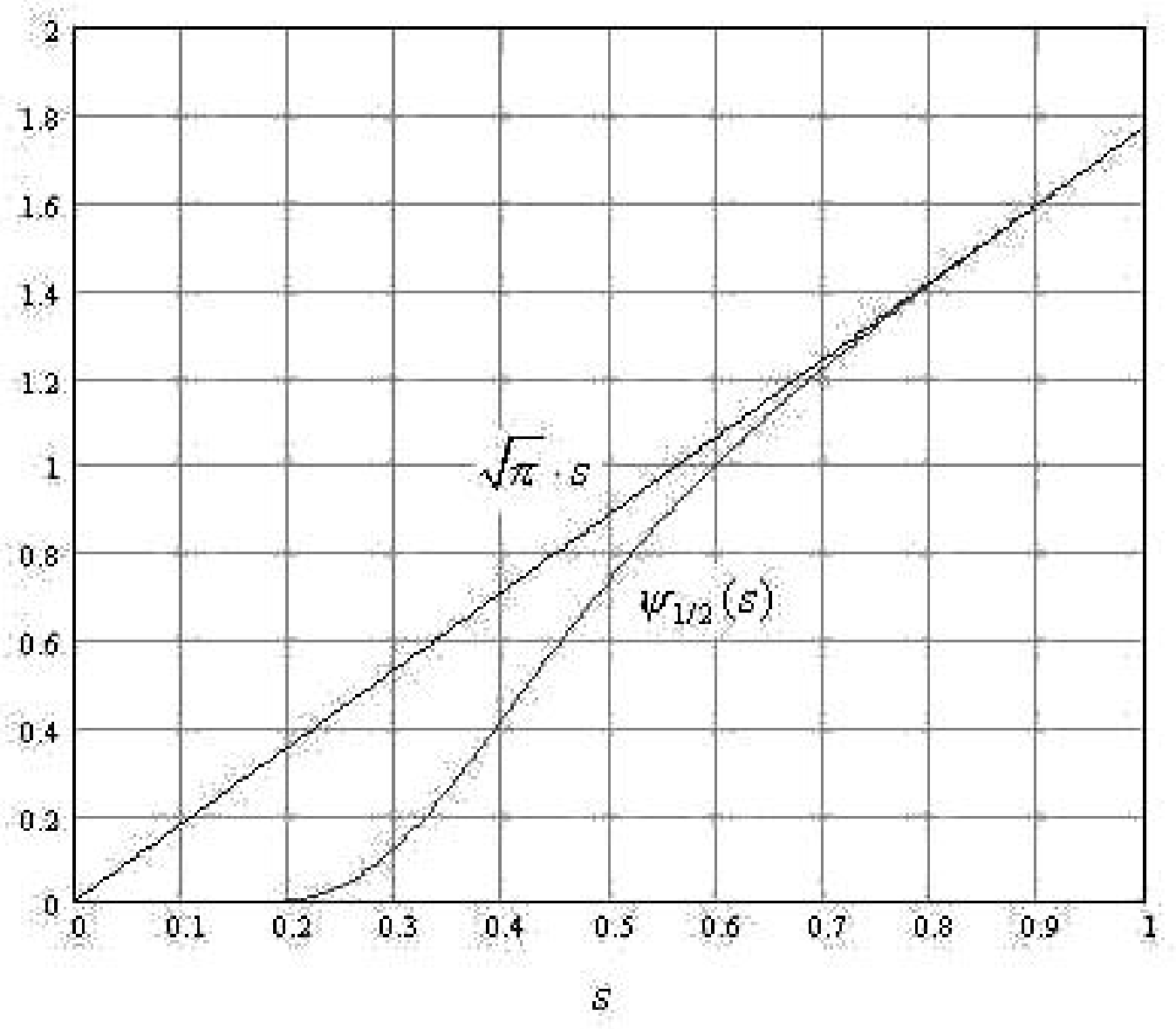

Fig.3. *Numerical calculus for* $y_{1/2}(s)$

## *Asupra functiei* $\sum_{k=-\infty}^{\infty} \exp(-\frac{k^2}{s^2})$

***Rezumat:*** *Sunt studiate numeric si analitic unele proprietãti ale functiei*

$y(s) = \sum_{k=-\infty}^{\infty} \exp(-\frac{k^2}{s^2})$ .

¶